\documentclass{icrc29}
\usepackage{graphicx,amssymb,amsmath,times}
\begin{document}
\title[Energy measurement by QUEST ...]{Cosmic Ray Energy Measurement with EAS
Cherenkov Light: Experiment QUEST and CORSIKA Simulation}
\author[E.E. Korosteleva et al.] {E.E. Korosteleva$^a$, 
L.A. Kuzmichev$^a$, V.V. Prosin$^a$,
        \newauthor
 EAS-TOP COLLABORATION$^b$,\\
        (a) Skobeltsyn Institute of Nuclear Physics, Moscow State University,
	Russia\\  
        (b) INFN, IFSI, Universita' Torino, Italy       }
\presenter{Presenter: V.V. Prosin (prosin@dec1.sinp.msu.ru), \  
rus-prosin-VV-abs1-he12-poster}

\maketitle

\begin{abstract}
                                                                      
 A new method of a primary cosmic particle energy measurement with the extensive
 air shower (EAS) technique has been developed by exploiting:  
 a) the joint analysis of the shower size, obtained by the EAS-TOP array, 
 and of the EAS Cherenkov light lateral distribution (LDF), obtained by 
 the QUEST array, and b) simulations based on the CORSIKA code. 
 The method is based on the strict correlation between the size/energy ratio and 
 the steepness of the Cherenkov light lateral distribution and has been compared 
 with a "classical" one based on the Cherenkov light flux at a fixed distance 
 (175 m) from the EAS core. The independence of the energy measurement both on
 the mass of primary particle and the hadronic interaction model used for the
 analysis is shown.  
 Based on this approach the experimental integral intensity of cosmic rays flux
 with energy more than $3\cdot 10^{15}$~eV is obtained with good systematic and
 statistical accuracy.

\end{abstract}

\section{Experiment QUEST and Simulations}

The QUEST experiment was developed to combine wide-angle atmospheric Cherenkov
light measurements with the charged particle EAS-TOP measurements 
(Gran Sasso, Italy, 2000 m a.s.l.)\cite{1}. The wide-angle Cherenkov light
detector was based upon five QUASAR-370 ($37$~cm diameter)
hemispheric photomultiplier tubes installed on five telescopes (average
pointing at direction $\theta=34^\circ$,
$\varphi=167^\circ$). 

The size $N_e$ and core position for every
shower has been extracted from EAS-TOP data. The reconstructed Cherenkov light
lateral distribution function (CLDF) has been obtained from the Cherenkov light flux
measured by each detector at the known distance from the axis.     
A new fitting function, suggested by us in ref.~\cite{1}, has been used to
derive two main parameters of the EAS CLDF for every recorded event:
the light flux at core distance of $175$~m $Q_{175}$ and the LDF steepness,
defined as the ratio of the fluxes at $100$ and $200$~m from the axis:
$P=Q(100)/Q(200)$. 


The energy measurement methods are based on analysis of artificial showers data,
simulated with CORSIKA code\cite{2,3}. 
The total sample of 400 events was simulated for primary energy $1, 2, 4$ and
$8$~PeV, and zenith angles $\theta$ from $24^\circ$ to $39^\circ$, 180 of them
for primary protons and 180 for iron nuclei using QGSJET\cite{4} model of
hadron interaction and 20 for protons and 20 for iron using SIBYLL\cite{5}
model.  
 To derive in the analysis of simulation the EAS size
$N_e$comparable to the experimental one, we have taken into account both
electrons and muons and used the experimental procedure of size reconstruction
with NKG fitting function. 
 
We obtained from these simulated data the dependences:

1) of the mean depth of EAS maximum $X_{max}$ on energy $E_0$, shape and
standard deviation of $X_{max}$ distribution separately for $p$ and $Fe$
primaries,

2) of $P$ on the linear distance to EAS maximum $H_{max}$ and
standard deviation of the $P$ distribution for fixed $H_{max}$,separately for $p$
and $Fe$, 

3) of the size $N_e$ on $P$ and $E_0$ and the standard deviation of the $N_e$
distribution fo fixed $P$ for $p$ and $Fe$,

4) of $Q_{175}$ on $E_0$ and the standard deviation of the $Q_{175}$
distribution at a fixed $E_0$ separately for $p$ and $Fe$. 

Using all these parametrizations and generating as base independent parameters:
the primary energy $E_0$ distributed as a power law spectrum, the depth of EAS
maximum $X_{\max}$ distributed as an asymmetric $\Gamma$-distribution, the
shower  
axis direction and core position we generate
and analyse hundreds thousands of artificial events.
Experimental errors in core position and $N_e$ are inserted in this procedure in
accordance with \cite{6}. 
The real array geometry and fluctuations of every Cherenkov light detector
response 
are taken into account. We call the described procedure "model of experiment". 
It is used for analysis of experimental errors, efficiency and
distributions of measured parameters for different assumptions on primary
composition.

To analyze the experimental data we use different parametrisations of
CORSIKA simulation for complex initial composition. 
Figure~1 shows the connection of LDF steepness $P$ with liner distance to
EAS maximum $H_{max}$ in [km].
The best parametrisation of this dependence 
is:  $H_{max} = 12.65 - 1.85P$, 
with standard deviation of $H_{max}$ distribution for fixed $P$:
$\sigma(H_{max}) = 0.3 km$, - which may characterize the theoretical
accuracy of the method. 


\begin{figure}[t]
\includegraphics*[width=0.5\textwidth,angle=0,clip]{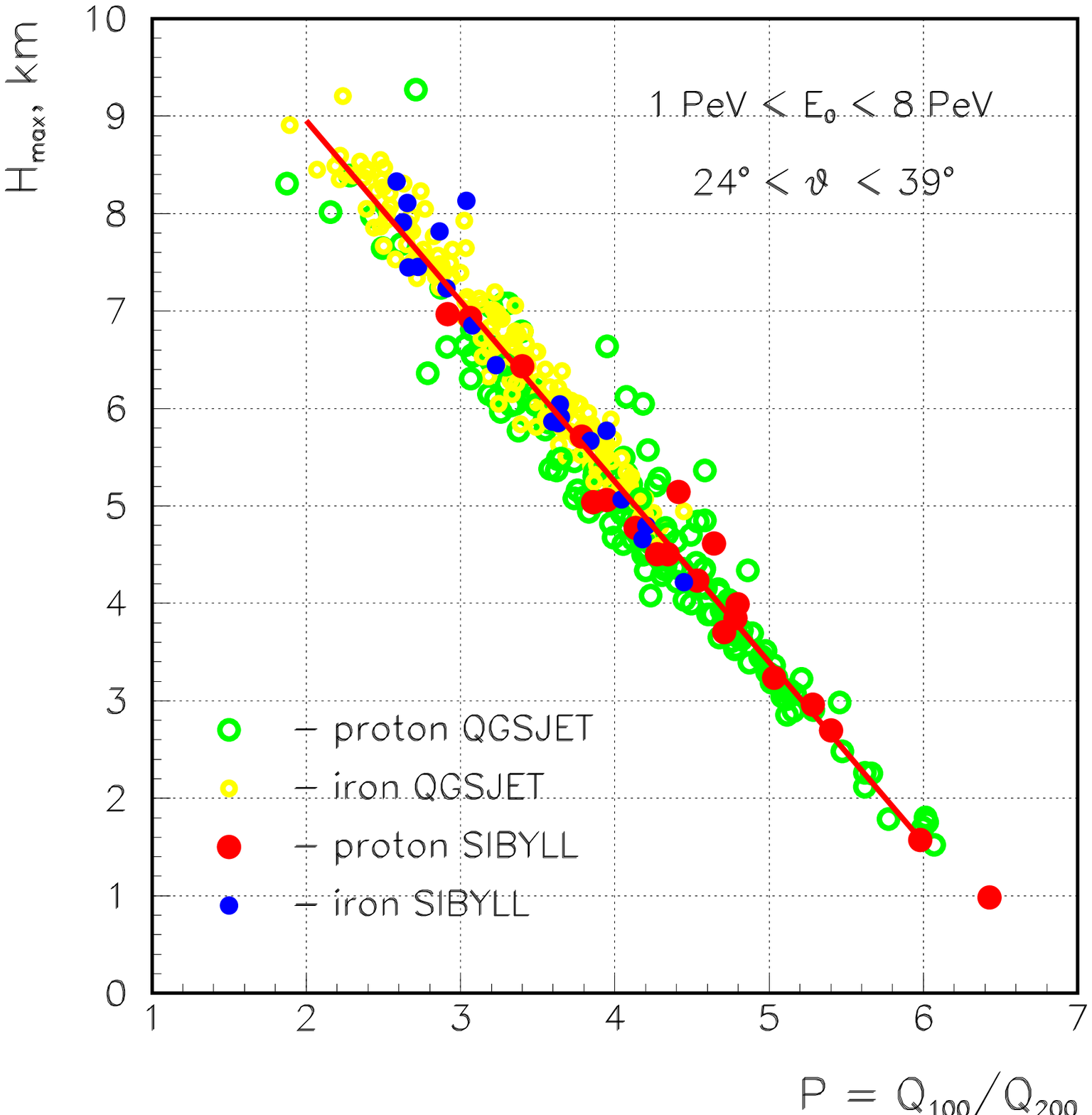}
\hfill
\includegraphics*[width=0.5\textwidth,angle=0,clip]{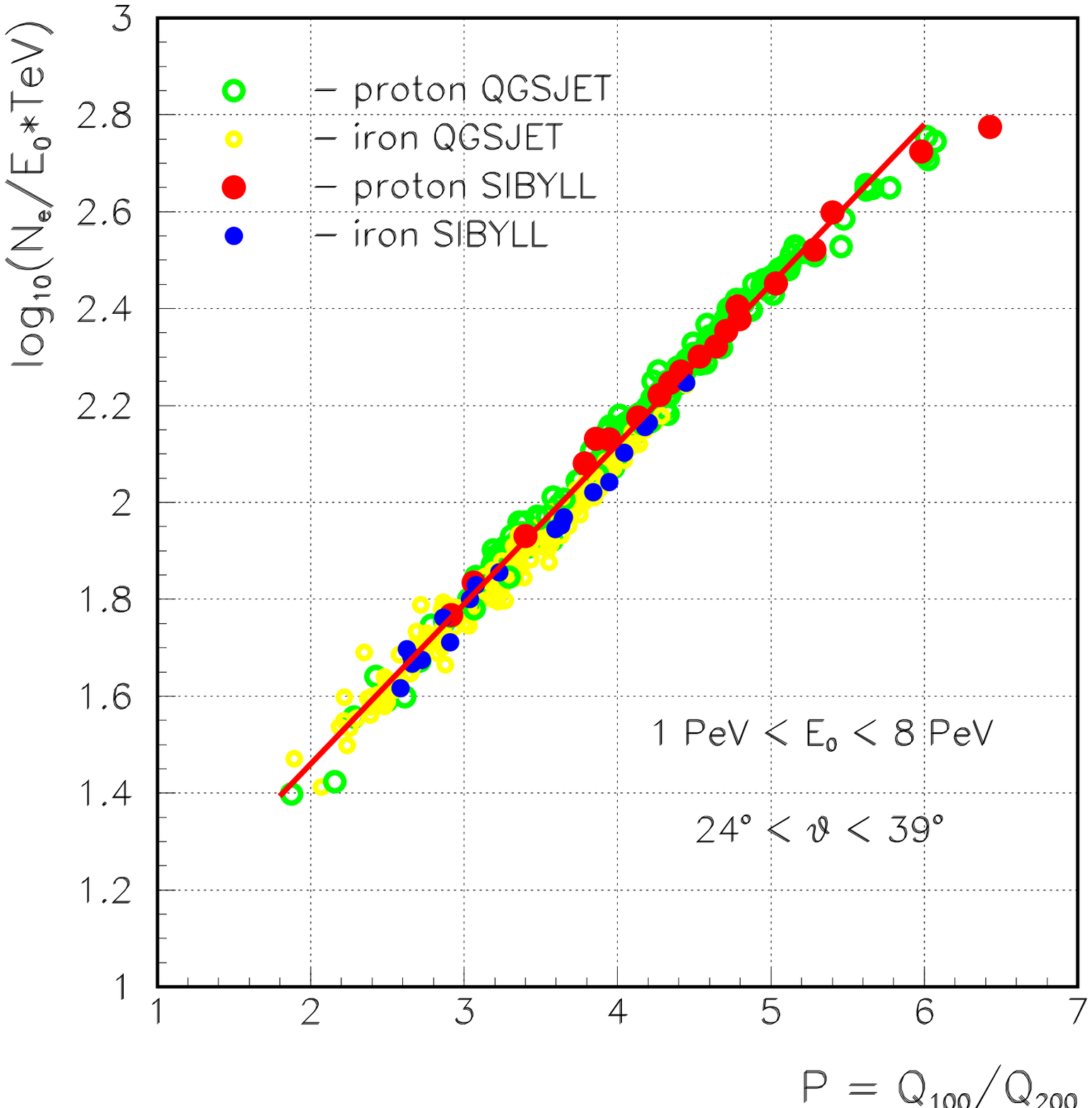}
\vspace*{8pt}
\parbox[t]{0.48\textwidth}{\caption{CORSIKA: 
Correlation between EAS Cherenkov LDF steepness $P$ and $H_{max}$.}}
\hfill
\parbox[t]{0.48\textwidth}{\caption{CORSIKA: 
Correlation between EAS Cherenkov LDF steepness $P$ and $N_e/E_0$.}}
\end{figure}


\section{SIZE/CLDF Method: Size and Cherenkov Light LDF Steepness P}

 Figure~2 shows the CORSIKA simulated correlation between the CLDF steepness $P$
and the ratio of the size to primary energy
($N_e/E_0$) for the 400 above discribed events. One can notice from fig. 1 and 2
that such relation is almost independent on
parameters: primary energy, zenith angle, sort of particle and
hadron interaction model. The correlation between $N_e/E_0$ and $P$ is more strict
than the one between $N_e/E_0$ and position of EAS maximum. 

The difference between $N_e/E_0$ for $p$ and $Fe$ primaries is less than 6\% and
for the two interaction models less than 2\% (for $P$=4). 
Using the correlation shown in fig.2 we can get the
primary energy in experiment from the measurement of $N_e$ and $P$:
\[
	E_{SIZE}\;[\mbox{eV}] = 1.59\times10^{11}\;N_e/exp(0.76P) \eqno (1)
\]

The main practical advantage of this method
relies in the well developed technique of scintillator response calibration
based on the measurement of the single particle response\cite{6}. 
However the "model of experiment", described above, gives an
error of individual measurement of about $35$\% ($\sigma(
log_{10}(E_{SIZE}/E_{0})) = 0.129$) mostly due to the experimental
error of parameter $P$.   
 
Similar method of energy reconstruction, but for LDF steepness, estimated at
smaller distances from the core ($20-100$~m), was suggested in Ref.~\cite{7}.

\section{$Q_{175}$ Method: Cherenkov Light Flux at $175$~m Core Distance}  

\begin{figure}[t]
\includegraphics*[width=0.5\textwidth,angle=0,clip]{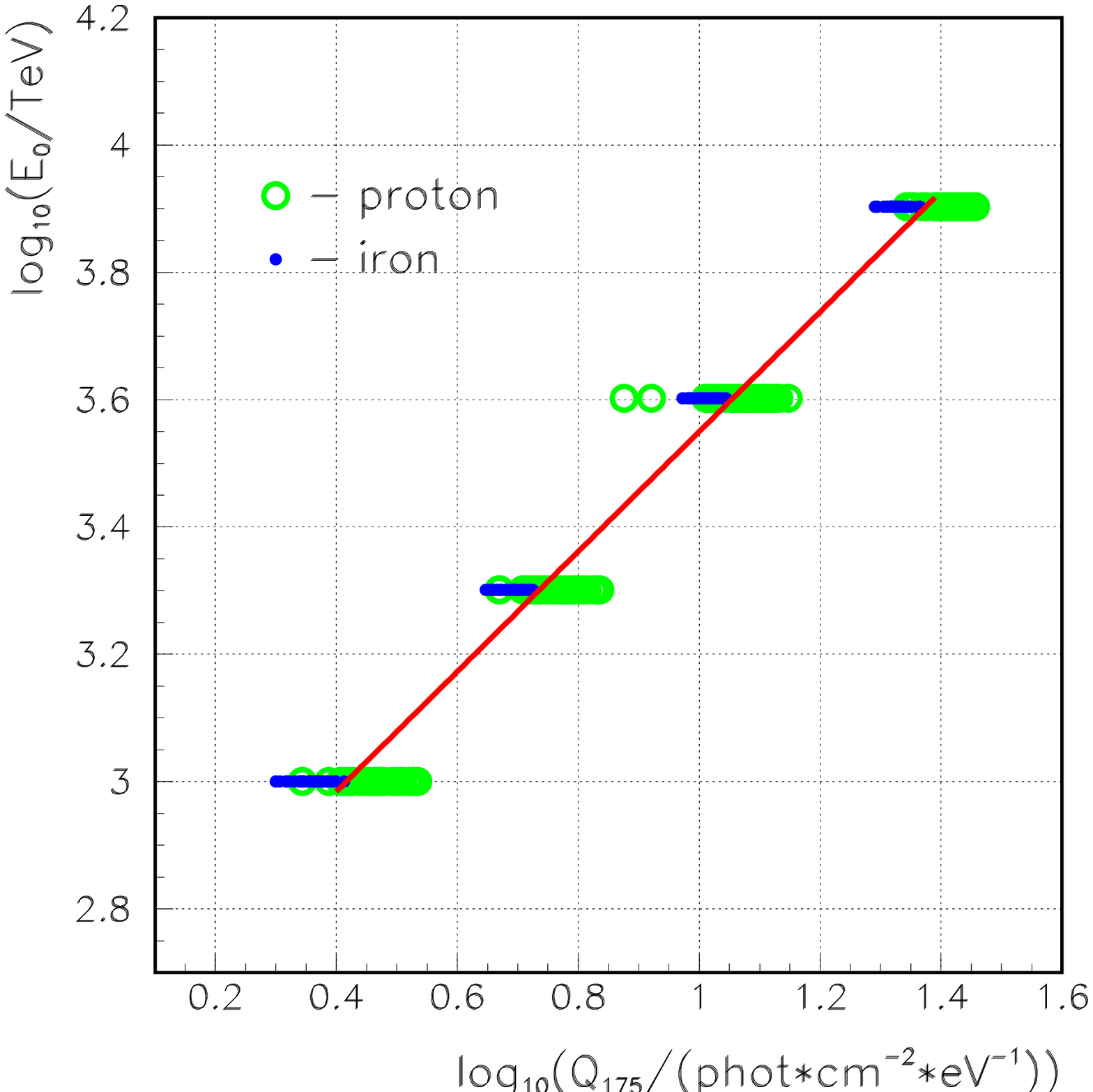}
\hfill
\includegraphics*[width=0.5\textwidth,angle=0,clip]{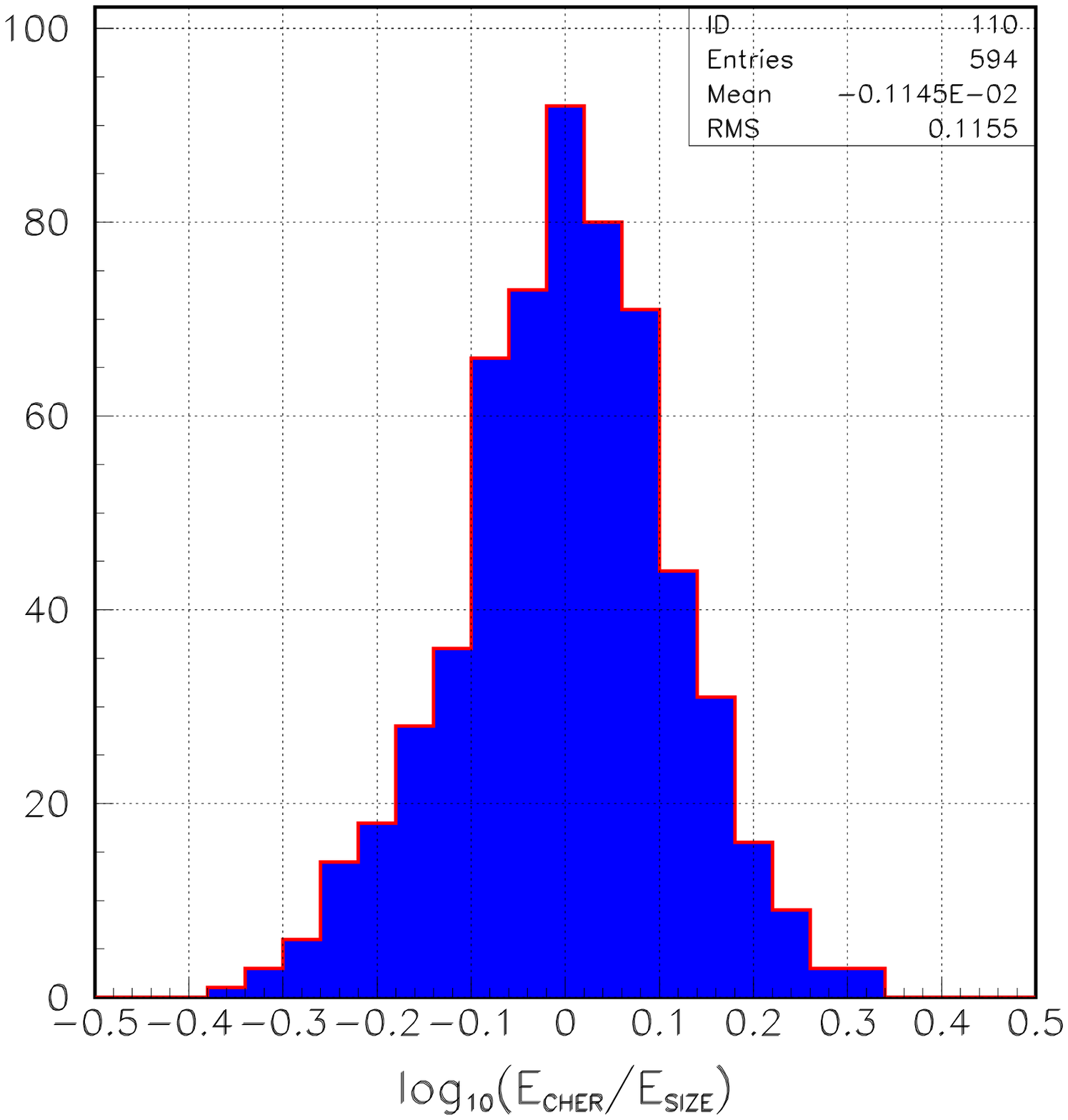}
\vspace*{8pt}
\parbox[t]{0.48\textwidth}{\caption{CORSIKA: 
Energy measurement by $Q_{175}$}}
\hfill
\parbox[t]{0.48\textwidth}{\caption{EXPERIMENT: Comparison of energy, 
obtained with two different methods for $E_{CHER}\geq 3\times10^{15}$~eV.}}
\end{figure}

Figure. 3 shows the CORSIKA simulated correlation between the primary energy and
the parameter $Q_{175}$. Taking into account the distribution of the 400 points
shown in 
fig. 3 we derive out an almost proportional relation between $E_0$ and $Q_{175}$: 
	$E_{CHER} = C\cdot Q_{175}^{0.94}$. 


The main problem of this "classical" method, used in many works, is in
the absolute calibration coefficient $C$, if one includes the systematic
uncertainty of $Q_{175}$ in it.
The error of absolute calibration has been estimated from  $18$\%
to $30$\% for different experiments. To get better accuracy we suggest to
use the mean experimental ratio $<E_{SIZE}/Q_{175}^{0.94}>$, as the
coefficient for the absolute calibration of Cherenkov array response. 
So finally:
$$
	E_{CHER} = <E_{SIZE}/Q_{175}^{0.94}>\cdot Q_{175}^{0.94} \eqno (2)
$$

"Model of experiment" displays an experimental uncertainty about $15$\% for 
the energy measurement by such expression 2, i.e. the accuracy of every
individual energy 
measurement is much better than for the SIZE/CLDF method.
 
The final comparison of experimental energy obtained with two methods is shown
in fig.~4.
The standard deviation of the experimental distribution is very
close to that obtained with the "model of experiment" for the
SIZE/CLDF method, that confirms indirectly the experimental errors
estimation of the "model of experiment".

\section{A Reference Integral Cosmic Rays Intensity}

The energy measured with the Cerenkov light flux method is used for
estimation of the integral intensity of cosmic rays, since the experimental
error of this method for individual event is at least 2 times smaller than
that for the SIZE/CLDF method. 
The systematic uncertainty in the definition of the integral intensity is
mainly due to the estimation of the
threshold energy. The main contribution to it is the uncertainty in 
the size $N_e$, which is evaluated as less than 6\%~\cite{6}.This leads to an
uncertaity of about 12\% in the integral intensity. 

The maximum systematic
shift of calibration coefficient, connected with the lack of knowlege of the real
mass composition, was
estimated with "model of experiment" assuming pure proton and pure iron
compositions. The maximum error of about 8\% is obtained for primary protons. We may
estimate the maximum possible systematic uncertainty as a root mean square of
the sum of squares of these two values.      

To analyze the experimental data we used the events with reconstructed core
positions inside 
the effective area of 100$\times$100~m$^2$ in the center of EAS-TOP array, zenith
angles less than 40$^\circ$ and relative angles of the axis to the Cherenkov
average array pointing less than 34$^\circ$. 100\%
efficiency for such events is reached at $2.5\cdot 10^{15}$~eV, as obtained with
the "model of experiment".  
     
$594$ events have been recorded during $140$~h of data taking with energy
larger than
$3\times10^{15}$~eV. The corresponding integral intensity is
\[
I\left(E_0\ge3\times10^{15}\;\mbox{eV}\right)=(2.3\pm0.1^{stat}\pm0.4^{syst})\times10^{-7},
 [\mbox{m}^{-2}\cdot\mbox{s}^{-1}\cdot\mbox{ster}^{-1}].
\]

A practical estimation of the stability of the integral intensity was obtained
by 
dividing the whole staistics into two parts, one of them acquised during summer
and autumn and another acquised  during winter and early spring. The
natural conditions of the experiment were quite different, but the difference in
estimation of the reference integral intensity is 8\% only, consistent with the
possible statistical error.

The obtained integral intensity can be used as a reference point for
other cocmic ray experiments having no precise absolute calibration.  




\end{document}